# Observation of subluminal twisted light in vacuum: comment

PEETER SAARI*

*Institute of Physics, University of Tartu, W. Ostwaldi 1, 50411, Tartu, Estonia*
*peeter.saari@ut.ee*

**Our analysis based on calculations carried out with different methods but with the data taken from [Optica, 3, 351 (2016)] lead to the conclusion that results obtained in [Optica, 3, 351 (2016)] are questionable in several respects.** © 2016 Optical Society of America

Bouchard *et al.* [1] showed recently that Laguerre-Gauss (LG) pulses exhibit subluminal group velocities in vacuum, being 0.1% slower relative to *c* in the focal region. Such a substantial subluminality effect is intriguing, since it has been known for a long time that fundamental (zeroth-order) Gaussian beam pulses exhibit a slight *super*luminality, i.e., they propagate a little faster than *c* through the focus [2]. The authors [1] stress the importance of the observed effect in applications involving "twisted" light pulses. Moreover, the subluminality resulting in well measurable retardation times (about 20 fs in a typical paraxial geometry) would be very prospective for various novel experiments with light pulses possessing orbital angular momentum.

However, the results of Ref. [1] seem to be incorrect in some respects. In what follows we analyze the theoretical evaluations of Ref. [1] with the help of formulas from Ref. [2] and compare the curves of the pulse's subluminal behavior obtained in [1] with those resulting from our calculations carried out in various ways, but with the same parameters as in Ref. [1]. Unfortunately, we reach the conclusion that the subluminality effect cannot be so large—the group velocity drop in the focal region should be more than by an order of magnitude smaller, and the velocity reduction as a function of the propagation distance should be more complicated. Finally, we discuss briefly the behaviors of group velocities of other types of Gaussian pulses.

The calculation of the group velocity in Ref. [1] starts from the expression for position- and frequency-dependent phase of the LG mode, viz.,

$$\Phi^{LG}(r, z, \varphi, \omega) = \frac{\omega}{c} z + \frac{\omega r^2}{2cR} + l\varphi - (2p + |l| + 1)\zeta, \quad (1)$$

where $R := R(z, \omega)$ is the radius of curvature of the beam phase front, $\zeta := \zeta(z, \omega)$ is the Gouy phase, and *p, l* are the indices of the LG mode. Subsequently, the authors use the well-known formula $v_g = |\partial_\omega \nabla \Phi|^{-1}$ for calculating the group velocity. Unfortunately, the resulting expression is not shown in Ref. [1], obviously because it is too long and complex. Yet the results of the calculations of the normalized group velocity change $(v_g - c)/c$ (henceforth NGVC) as curves depending on propagation distance and mode indices are presented in Fig. 1 of Ref. [1]. Fig. 1(b) reveals the essence of the findings of the authors: the NGVC curves as functions of the axial coordinate *z* exhibit a negative peak at the focal region, which deepens linearly with the index *l* (the index *p* has been set to zero, the same is done in this comment). This is in contrast with the fundamental Gaussian pulse (i.e., when $l = 0$) whose NGVC curve has a positive peak (i.e., it is superluminal) within the Rayleigh range. If the group velocity of an LG pulse with $l > 0$ was measured on the propagation axis *z*, it would turn out to be superluminal around the focus as well. But since the intensity of an LG pulse is concentrated on a ring whose radius is proportional to $l^{1/2}$, one has to consider the group velocity not at points on the axis *z* but outside of it—on the ring. Exactly that was done by the authors of Ref. [1] and results of their calculations in Fig. 1(b) show the subluminality effect.

We do not question the effect and its linear dependence on the index *l*. Indeed, the linear dependence of the NGVC can be inferred already from an inspection of Eq. (1). Namely, after taking derivatives with respect to *ω* and *z* according to the expression $v_g = |\partial_\omega \nabla \Phi|^{-1}$, the term $l\varphi$ disappears and the first term reduces to $c^{-1}$. Further, the last term with the Gouy phase is explicitly linear with respect to *l* and gives a superluminal contribution to the NGVC. The second term is also linear (since the radius $r_{max}$ of the ring of the maximum intensity of a LG mode is proportional to $\sqrt{l}$) and is responsible for the competing subluminal contribution. Near the focus, all derivative terms originating from these two are much smaller than $c^{-1}$ and, therefore, taking the modulus and reciprocal value does not practically affect the linearity, and the NGVC turns out, indeed, to be proportional to the index *l*.

What we question is the shape of the dependence of the group velocity reduction on the propagation distance *z*, which in Ref. [1] has been found to comprise a single negative peak at $z = 0$ [see Fig. 1(b) in Ref. [1]]. If we plot the expression for the NGVC derived analytically by us and verified (for the case of $l = 0$) with the group velocity formulas in Ref. [2], we obtain a curve with two minima and a *positive* peak in the center, as depicted in Fig. 1 here.

If we express the waist of the focused beam in units of the Rayleigh range $z_R$ as $\overline{w_0} = w_0/2z_R$, then our expression for the NGVC at point $(z = 0, r = r_{max})$ reduces to $(2 - |l|)\overline{w_0}^2$ and at points $(z = \pm z_R, r = r_{max})$, it reduces to $(-|l|)\overline{w_0}^2$. From these simple expressions, one can draw the following conclusions: (i) the NGVC curve has two minimums and a maximum in the center irrespective of the parameters of the beam (if $l > 0$); (ii) appearance of the difference $2 - |l|$ in the expression for $z = 0$ is logical due to the two aforementioned small terms of opposite sign in Eq. (1); (iii) in the case of $l = 2$ the subluminal and superluminal contributions into the group velocity compensate each other at $z = 0$; and (iv) since for any paraxial geometry, the condition $\overline{w_0^2} \ll 1$ is very well fulfilled, the higher-order (nonlinear) terms are negligible in the expressions for the group velocity. For the beam parameters taken from the caption of Fig. 1(b) we obtain $\overline{w_0^2} = 9.8 \times 10^{-6}$, which is about 16 times smaller than in the expression $-1.6 \times 10^{-4} l$ given for the NGVC in Ref. [1]. Moreover, the proportionality factor $-1.6 \times 10^{-4}$ is said to be calculated from the beam waist $w_0 = 100\ \mu m$ and with this value the discrepancy is 100-fold, as $\overline{w_0^2} = (2\pi w_0/\lambda)^{-2} = 1.6 \times 10^{-6}$. The same small value follows from Eq. (14) of Ref. [3], where an averaged group velocity has been calculated. Hence, we also question the strength of the group velocity reduction effect found in Ref. [1].

At this point, we must admit that our analytical derivations were carried out—like in Ref. [2]—under the assumption that a known small shift in position between the geometrical focus and the LG beam waist is negligible. This assumption not only is well justified in the given paraxial geometry—the convergence angle of the focused beam is as small as 0.36 deg—but also avoids too-cumbersome expressions in the derivations.

Nevertheless, for a strict comparison, we carried out a numerical computation of the NGVC following the procedure described in the Supplementary material of Ref. [1], i.e., taking into account Eq. (S11) which expresses the radius $R$ of the phase front curvature without the above-mentioned assumption. Our result for $l = 5$ is shown in Fig. 1. The curve is indistinguishable from the one obtained by our analytical expression derived under the above-mentioned assumption. This is because the factor $1/R(z, \omega)$ in Eq. (1) differs less than 0.1% if calculated with and without the assumption.

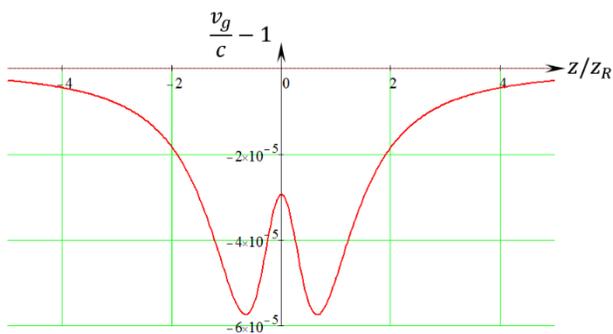

**Fig. 1**. Group velocity at the radial position corresponding to the LG beam (with $l = 5, p = 0$) intensity maximum as a function of propagation distance. The curve has been calculated numerically with the same beam parameters as in Fig. 1(b) of Ref. [1] (pulse carrier wavelength 795 nm, beam waist 2.5 mm in the plane of a thin lens of focal length 400 mm), but the curve substantially differs from the corresponding one in Fig. 1(b) of Ref. [1]. At the same time, it coincides—as it should—with the curve plotted from our analytical expression obtained under the very well fulfilled assumption considered in the text.

The two minimums in Fig. 1 merge into one (like in Ref. [1]) if we calculate the curve for such radial distances that exceed more than 1.5 times the radial distance of the beam maximum. But such curves are senseless since at such radial positions, the beam intensity is less than 10% of the maximum value. Comparing Fig. 1 here and in Ref. [1] quantitatively, we see that according to our calculations the group velocity reduction effect is almost 20 times weaker. This means that the delays caused by the subluminality of the pulse should be of the order of 1 fs instead of tens of fs as reported in Ref. [1], which would require for their measurement an elaborated sub-femtosecond technique

The experiments carried out by the authors of Ref. [1] are in accordance with their theoretical results: measured delays grow linearly with $l = 0, 1, 2, \ldots 6$ up to 26 fs. The autocorrelation traces for obtaining the delays were measured with a surprisingly high SNR, and the authors were able to observe few-micrometer shifts of $201 \pm 6\ \mu m$ wide autocorrelation traces, which is surprising also. We do not want—and one cannot without knowing all the relevant technical details—to comment further on their experiments. However, a question arises whether the grating patterns on the SLM could not contribute into the $l$-dependence of the delays, since it is known that diffractive optical elements cause observable subluminal propagation of light pulses (see [4] and the references therein).

As a matter of fact, when calculating the delay experienced by the LG pulse ($l > 0, p = 0$) propagating over a distance along the axis $z$, one must not use the absolute value of the group velocity as done in [1], but, instead, e. g., its projection onto the axis $z$. Such quantity does not reach $c$ far from the focus but, instead, a constant value $v_{gz} \approx c(1 - 1 \times 10^{-5} l)$ for the given beam parameters, resulting in the delay about 70 fs over one focal length if $l = 5$. A similar delay results from Eq.(3) of [3] which gives an averaged $v_{gz}$. Hence, the comparatively large experimental values of the delay up to 26 fs measured in [1] need not to be incorrect.

Finally, it is appropriate here to consider other possibilities to generate subluminal LG pulses. Pulses built from a given monochromatic mode are of various types, with different propagation properties, determined by how the (interrelated) parameters of the mode as a Fourier constituent of the pulse depend on the frequency or wavelength [5]. For example, frequency-independent Rayleigh range results in the so-called isodiffracting Gaussian beam pulse [6], whose peak propagates with strictly luminal group velocity. In this way one can distinguish three types of Gaussian pulses [7] and even four types of Airy pulses [8, 9]. Contrary to the case of frequency-independent input waist—just the case considered so far, when the on-axis group velocity is superluminal in the focal region—in the case of frequency-independent waist of the focused beam the group velocity is *subluminal* even on the axis [7, 10]. Parenthetically, let us note that superluminal group velocity of the pulses considered in Ref. [1] and here does not allow superluminal signaling since the waist $w_0$ is inversely proportional to frequency and, therefore, the pulse transforms substantially its shape in the focal region. The last circumstance, by the way, may affect the autocorrelation measurements.

In the case of frequency-independent $w_0$, the Gouy phase term in Eq. (1) gives also a subluminal contribution to the group velocity, so that the positive peak in Fig. 1 should disappear. Unfortunately, our calculations showed that this contribution results in less than a two-fold reduction of the group velocity in comparison to Fig. 1.

In conclusion, our analysis and calculations force us to question in several respects the interesting and promising results obtained in Ref. [1].

**Funding.** Eesti Teadusagentuur (Estonian Science Foundation) (PUT369).

**Acknowledgment**. The author is thankful to Yannis Besieris for valuable comments.